\documentclass[prd,aps,twocolumn,nofootinbib]{revtex4}

\usepackage{graphicx}

\begin{document}

\title{511 keV line and diffuse gamma rays from moduli}

\author{Shinta Kasuya$^a$ and Masahiro Kawasaki$^{b}$}

\affiliation{
$^a$ Department of Information Science,
     Kanagawa University, Kanagawa 259-1293, Japan\\
$^b$ Institute for Cosmic Ray Research,
     University of Tokyo, Chiba 277-8582, Japan}

\date{February 14, 2006}

\begin{abstract}
We obtain the spectrum of gamma ray emissions from the moduli whose decay into $e^+ e^-$
accounts for the 511 keV line observed by SPI/INTERGRAL. The moduli emit gamma rays
through internal bremsstrahlung, and also decay directly into two gammas via tree and/or 
one-loop diagrams. We show that the internal bremsstahlung constrains the mass of the moduli 
below $\sim 40$ MeV model-independently.  On the other hand, the flux of two gammas directly 
decayed from the moduli through one loop diagrams will exceed the observed galactic diffuse
gamma-ray background if the moduli mass exceeds $\sim 20$ MeV in the typical situation. 
Moreover, forthcoming analysis of SPI data in the range of $1-8$ MeV may detect the line emisson with the energy half the moduli mass in the near future, which confirms the decaying
moduli scenario.
\end{abstract}


\maketitle

\setcounter{footnote}{1}
\renewcommand{\thefootnote}{\fnsymbol{footnote}}

\section{Introduction}
The spectrometer SPI on the International Gamma-Ray Astrophysics Laboratory (INTEGRAL) 
confirmed with its high resolution that 511 keV line gamma-rays  are coming from the Galactic 
bulge \cite{SPI}. It is explained by the annihilation of positrons with electrons, but the source 
of the positrons is difficult to be traditional astrophysical objects, since the emission 
region observed is wide-spread. Thus, it may be plausible that the particle dark matter 
annihilating or decaying into positrons could explain the observed 511 keV line 
\cite{DDM1,KY05,DDM2}. 

Among them, the moduli may be the simplest candidate, since it appears very naturally in
supersymmetric and/or superstring models \cite{KY05}. One of the prominent feature of the
moduli is that they can decay into two gammas directly via tree or one loop diagrams. This could
be a signal for proving the scenario once one would observe the line emission with the energy
half of their mass, i.e., $m_{\phi}/2$, where $m_{\phi}$ is the moduli mass. 

In this article, we consider the photon emission from the moduli through the internal 
bremsstahlung \cite{Beacom1}, and direct productions via tree or one-loop diagrams in our 
galaxy, as well as from extragalactic emission. We evaluate these line and/or diffuse 
gamma-ray spectra, assuming that the flux of the 511 keV line in the galactic center, 
observed by SPI/INTEGRAL, is explained by the positron production by the moduli decay. 
Comparing them with the diffuse gamma-ray backgound observations from INTEGRAL, 
COMPTEL, EGRET, and SMM, we obtain the constraints on the moduli mass.

\section{511 keV line}
We consider the moduli $\phi$ couple to electrons by the following Yukawa coupling \cite{KY05}:
\begin{equation}
\label{yukawa}
{\cal L} = \frac{m_e}{M_P} \phi \bar{e} e,
\end{equation}
where $m_e$ is the electron mass, and $M_P \simeq 2.4 \times 10^{18}$ GeV is the (reduced)
Planck mass. The decay width is given by
\begin{equation}
\Gamma_e \equiv \Gamma(\phi\rightarrow e^+ + e^-) \simeq 
\frac{1}{8\pi} \left(\frac{m_e}{M_P}\right)^2 m_\phi.
\label{tree1}
\end{equation}
In order to explain the 511 keV line gamma-rays from the galactic center with the flux of 
$\Phi_{511} \simeq 10^{-3}$ cm$^{-1}$ sec$^{-1}$ observed by SPI/INTEGRAL \cite{SPI},
the amount of the moduli should be \cite{DDM1,KY05}
\begin{equation}
\Omega_{\phi} \simeq \left(\frac{\Phi_{511}}{10^{-3} {\rm cm}^{-1} {\rm sec}^{-1}}\right)
\left(\frac{\tau_{\phi}}{10^{27} {\rm sec}}\right) \left(\frac{m_\phi}{\rm MeV}\right),
\end{equation}
where $\Omega_{\phi}$ and $\tau_\phi$ is the density parameter and the lifetime of the moduli,
respectively, and we apply a spherically symmetric profile for the dark matter density with
$\rho\propto r^{-1.2}$. Taking $\tau_{\phi} = \Gamma_e^{-1}$, the abundance should thus be
$\Omega_\phi \simeq 4\times 10^{-4}$. \footnote{
Moduli could contribute considerably to dark matter, i.e., $\Omega_\phi \sim \Omega_{DM} \sim
0.3$, provided that the coupling of the moduli to electrons is 
${\cal L} = s \frac{m_e}{M_P} \phi \bar{e} e$  with $s \sim 0.03$. Notice that this change 
does not alter the constraints of internal bremsstahlung or decay into two photons discussed
in the following sections.}

\section{Internal bremsstahlung}
\begin{figure}[!h]
\includegraphics[width=80mm]{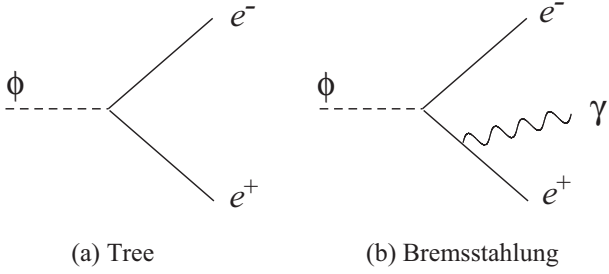}
\caption{Diagrams of decay of the moduli into positron and electron pair with (b) and without (a)
internal bremsstahlung photon.
\label{fig_diagram1}}
\end{figure}

Let us start with investigating the internal bremsstahlung spectrum. In the context of annihilating
dark matter, the spectrum of the gamma-ray due to the internal bremsstahlung has been obtained 
in Ref.~\cite{Beacom1}, and it is found that the energy of the emitted photons should be smaller
than $\sim 20$ MeV. Here we consider analogous process in the context of decaying particles.
The tree diagram of the decay process is shown in Fig.~\ref{fig_diagram1}(a). The internal
bremsstahlung photons are emitted as in Fig.\ref{fig_diagram1}(b). The decay rate for the internal
bremsstahlung is written as (See the Appendix)
\begin{eqnarray}
& & \frac{d\Gamma_{br} }{d\varepsilon_\gamma} = \frac{y^2 e^2}{(2\pi)^3} \nonumber \\
& & \times \left[ \left(
\frac{\varepsilon_\gamma}{m_\phi}-1+\frac{m_\phi}{2\varepsilon_\gamma}
-\frac{2m_e^2}{m_\phi \varepsilon_\gamma}+\frac{4m_e^2}{m_\phi^2}\right)
\log \frac{1+\eta}{1-\eta} \right.  \nonumber \\
& & \qquad\qquad \left. +\left( \frac{m_e^2}{m_\phi \varepsilon_\gamma} 
+\frac{2m_e^4}{m_\phi^3\varepsilon_\gamma} \right) \frac{2\eta}{1-\eta^2} \right], 
\end{eqnarray}
where $y=m_e/M_P$, $\varepsilon_\gamma$ is the photon energy and
\begin{equation}
\eta=\left(1-\frac{4m_e^2}{m_\phi(m_\phi-2\varepsilon_\gamma)}\right)^{1/2}.
\end{equation}

Since the (tree) decay rate into positron and electron pair is given by
\begin{equation}
\Gamma_e =\frac{y^2 m_\phi}{8\pi} \left(1-\frac{4m_e^2}{m_\phi^2}\right)^{3/2},
\label{tree2}
\end{equation}
the flux spectrum for the internal bremsstahlung, normalized to the observed 511 keV line flux, 
becomes~\cite{Beacom1}
\begin{eqnarray}
\frac{d \Phi_{br}}{d\varepsilon_\gamma} &   \simeq & \frac{1}{2} \Phi_{511} \frac{1}{\Delta\Omega}
\left[ \frac{f}{4} +(1-f) \right]^{-1} \frac{1}{2} \frac{1}{\Gamma_e} 
\frac{d\Gamma_{br}}{d\varepsilon_\gamma} \nonumber \\
& \simeq & \frac{1}{4} \Phi_{511} \frac{1}{\Delta\Omega} \left[ \frac{f}{4} +(1-f) \right]^{-1}
\nonumber  \\
& & \times \frac{4\alpha_{em}}{\pi m_\phi} \left(1-\frac{4m_e^2}{m_\phi^2}\right)^{-3/2}
\nonumber \\
& & \times \left[ \left(
\frac{\varepsilon_\gamma}{m_\phi}-1+\frac{m_\phi}{2\varepsilon_\gamma}
-\frac{2m_e^2}{m_\phi \varepsilon_\gamma}+\frac{4m_e^2}{m_\phi^2}\right)
\log \frac{1+\eta}{1-\eta} \right.  \nonumber \\
& & \qquad\qquad \left. +\left( \frac{m_e^2}{m_\phi \varepsilon_\gamma} 
+\frac{2m_e^4}{m_\phi^3\varepsilon_\gamma} \right) \frac{2\eta}{1-\eta^2} \right].
\end{eqnarray}
Here a half of the 511 keV flux is emitted from an angular region $\Delta\Omega$ given by the
Gaussian FWHM of 8$^\circ$, and $f$ is the fraction of the positron decay via positronium
formation over all the incident positrons \cite{BL}. 

\begin{figure}[!h]
\includegraphics[width=80mm]{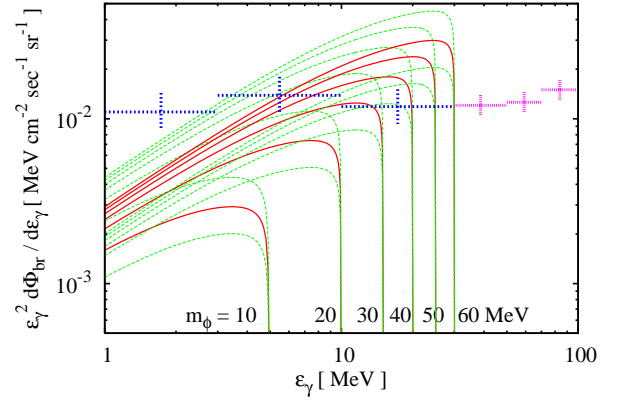}
\caption{Internal bremsstahlung spectrum for the various moduli masses. Dotted lines show
the uncertainty bands. Also plotted are the COMPTEL ($\varepsilon_\gamma <30$ MeV) and 
EGRET data ($\varepsilon_\gamma > 30$ MeV) of the galactic diffuse background.
\label{fig_brems}}
\end{figure}

We plot the spectrum for $m_\phi = 10 - 60$ MeV in Fig.~\ref{fig_brems}. The dotted lines 
represent the uncertainty coming from the observational uncertainties of 511 keV flux of
$\Phi_{511}=(1.07\pm 0.03)\times 10^{-3} {\rm cm}^{-1} {\rm sec}^{-1}$ and
$f=0.967\pm0.022$ \cite{SPI}. We also display the COMPTEL and EGRET data \cite{Strong}. 
In order not to exceed the observed flux, the mass of the moduli should be 
$m_\phi \lesssim 40$ MeV. Notice that this constraint is model independent, since  no
coupling constant $y$ appears in the formula of the spectrum. 

\section{Decay into two photons}
The moduli may decay into two photons through nonrenormalizable interactions suppressed by 
the gravitational strength. When the moduli are the dilation-type, they may decay via the following
interaction \cite{AHKY}:
\begin{equation}
{\cal L} = \frac{b}{4M_P}\phi F_{\mu\nu}F^{\mu\nu},
\end{equation}
where $b$ is of the order $O(1)$ parameter, which depends on the type of superstring theory
and compactification. Then the decay width via this interaction is given by
\begin{equation}
\label{dilaton}
\Gamma_{2\gamma}^{\rm tree} = \frac{b^2}{64\pi} \frac{m_\phi^3}{M_P^2}.
\end{equation}
Even the above interaction is absent, the moduli can decay into two photons though one loop
diagrams, since, at least, the moduli couple to positrons and electrons via Yukawa interaction
(\ref{yukawa}). In this case, the decay width is estimated as
\begin{equation}
\Gamma_{2\gamma}^{\rm loop} \simeq \frac{N^2}{4\pi} \left(\frac{\alpha_{em}}{4\pi}\right)^2 
\frac{m_\phi^3}{M_P^2}.
\label{loop}
\end{equation}
Here we include $N$, the number of the particle species running the loop, in the formula. Since
the moduli have to couple to electrons (positrons), it is likely to couple to other leptons, such as
muons and taus. Therefore, it may typically be $N=3$, but larger $N$ could be possible as well.

Let us first estimate the extragalactic diffuse gamma ray flux by the moduli decay. Since the
velocity dispersion of the moduli is negligible, two photons emitted have monochromatic energy
of $m_\phi/2$ when they are produced. Taking into account the dilution by the cosmic expansion,
the photon flux can be written as \cite{AHKY}
\begin{eqnarray}
& & \frac{d\Phi_{dif}}{d\varepsilon_\gamma} =  \frac{2 n_{\phi,0}\Gamma_{2\gamma}}{4\pi H_0}
\left(\frac{\varepsilon_\gamma}{m_\phi}\right)^{3/2} F(m_\phi/2\varepsilon_\gamma) \nonumber \\
& & \qquad \times \exp\left[ -\int_0^{m_\phi/2\varepsilon_\gamma} \!dx\frac{\Gamma_{2\gamma}}{H_0}
x^{-5/2} F(x) \right],
\label{diffuse_flux}
\end{eqnarray}
where $n_{\phi,0}$ is the present number density of moduli, 
$H_0 = 100h$~km/sec/Mpc is the present Hubble parameter and 
$F(x)=[\Omega_m + (1-\Omega_m-\Omega_\Lambda)/x+\Omega_\Lambda/x^3]^{-1/2}$.
Here we use $\Omega_m=0.27$, $\Omega_\Lambda=0.73$, and $h=0.71$~\cite{WMAP}. Since 
$\Gamma_{2\gamma}^{-1} \gg H_0^{-1}$, Eq.(\ref{diffuse_flux}) can be simplify to
\begin{equation}
\frac{d\Phi_{dif}}{d\varepsilon_\gamma}  = 
\frac{2 \Omega_\phi \rho_{C,0}\Gamma_{2\gamma}}{4\pi m_\phi  H_0}
\left(\frac{\varepsilon_\gamma}{m_\phi}\right)^{3/2} F(m_\phi/2\varepsilon_\gamma).
\label{extra-spec}
\end{equation}

\begin{figure}[!h]
\includegraphics[width=80mm]{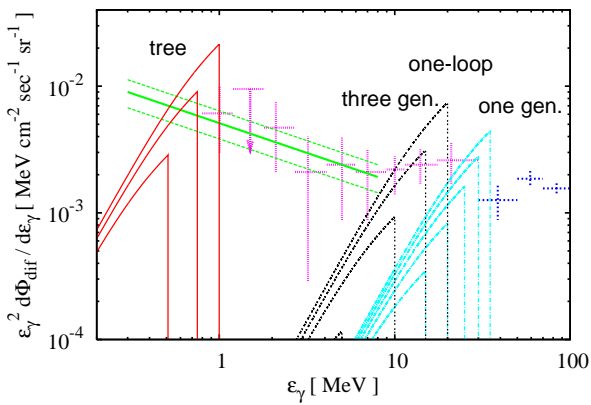}
\caption{Extragalactic diffuse gamma spectrum. We show $m_\phi=1.022$, 1.5, and 2 MeV 
for the direct decay rate (\ref{dilaton}), $m_\phi=10$, 20, 30, and 40 MeV for the one-loop decay
with $N=3$, and $m_\phi=30$, 40, 50, 60, and 70 MeV for $N=1$. Crosses are from 
COMPTEL  below 30 MeV, and EGRET data above 30 MeV. We also plot SMM data with thin
solid line with two dashed lines above and below, representing 25\% uncertainties.
\label{fig_dgb}}
\end{figure}

We plot the extragalactic diffuse gamma-ray spectra of three typical decay rates for various 
moduli masses in Fig.\ref{fig_dgb}. The observational constraints are from COMPTEL (below
30 MeV) \cite{Weidenspointer} and EGRET (above 30 MeV) data \cite{EGRET}. We also plot 
the data from the Solar Maximum Mission (SMM) Gamma-Ray Spectrometer in the energy 
range $0.3 - 8.0$ MeV \cite{Watanabe}. For the direct decay rate, the mass of the moduli is
very restricted such that $m_\phi \lesssim 1.5$ MeV. On the other hand, when two photons
are emitted through loop diagrams, the constraint on the mass is relatively less stringent:
$m_\phi \lesssim 30$ (60) MeV for $N=3$ (1).

The most stringent constraints will come from the two photon emission from our galaxy. Since 
the cosmological redshift is negligible, it results in a line spectrum with energy half the moduli 
mass, $m_\phi/2$. Since our assumption is that the annihilations of the positron from the moduli 
decay account for the observed 511 keV line in the galactic bulge, the ratio of the flux of 511 
keV line and two-gamma directly decayed from the moduli can be expressed as
\begin{equation}
\frac{\Phi_{2\gamma}}{\Phi_{511}}\simeq \frac{\Gamma_{2\gamma}}{\Gamma_e}
\left[\frac{f}{4}+(1-f)\right]^{-1}.
\end{equation}
Thus, the spectrum of the line gamma is given by
\begin{equation}
\varepsilon_\gamma^2 \frac{d\Phi_{2\gamma}}{d\varepsilon_\gamma} \simeq 
\frac{r \Phi_{2\gamma}}{\Delta\Omega}
\left(\frac{\varepsilon_\gamma}{\Delta \varepsilon_\gamma}\right)\varepsilon_\gamma,
\end{equation}
where $\Delta\Omega$ is the solid angle for $330^\circ < \ell < 30^\circ$ and $|b| < 10^\circ$,
and $r \approx 0.993$ is the fraction of the flux coming from the corresponding solid angle.
$\Delta\varepsilon_\gamma/\varepsilon_\gamma$ is the fractional energy resolution of the 
observation. We adopt $\Delta\varepsilon_\gamma/\varepsilon_\gamma \simeq 0.08$ for 
$\varepsilon_\gamma=1-30$ MeV for COMPTEL, while
 $\Delta\varepsilon_\gamma/\varepsilon_\gamma \simeq 0.13$ for 
 $\varepsilon_\gamma = 30 -100$ MeV for EGRET.  For INTEGRAL, the energy resolution is
 $\Delta \varepsilon_\gamma \simeq 2$ keV for $\varepsilon_\gamma = 0.5 -1$ MeV. 
 
Let us first examine the case which the moduli decay into two photons by the tree decay
width (\ref{dilaton}). Then the ratio of the flux becomes
\begin{equation}
\frac{\Phi_{2\gamma}}{\Phi_{511}}\simeq \frac{b^2}{8} \left(\frac{m_\phi}{m_e}\right)^2
\left[\frac{f}{4}+(1-f)\right]^{-1}.
\end{equation}
Since $[f/4+(1-f)]^{-1}\simeq 3.6$, the two-gamma flux is much larger than 511 keV line, which
is completely excluded by observations. For $m_\phi=2m_e$, both of the component could
explain the 511 keV line, if we lower the moduli abundance. In this case, however, continuum
spectrum just below 511 keV cannot be explained by $3\gamma$ continuum flux from
positronium formation, which was the confirmation of the 511 keV line by positron annihilation 
\cite{Strong2}.

\begin{figure}[!h]
\includegraphics[width=80mm]{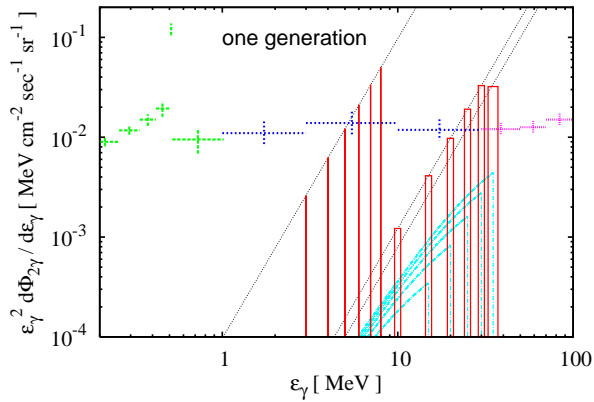}
\caption{Line spectra of two photons decayed via one-loop diagram with $N=1$ from the 
moduli in our galaxy. Three dotted lines show the fractional energy resolution of 0.13, 0.08, and 
0.001, from the bottom to the top. Data are from INTEGRAL ($\varepsilon_\gamma < 1$ MeV)
\cite{Strong2}, COMPTEL ($\varepsilon_\gamma = 1 - 30$ MeV), and EGRET ($\varepsilon_\gamma > 30$ MeV) \cite{Strong}. Extragalactic diffuse emission spectra 
(\ref{extra-spec}) are also shown for comparison.
\label{fig_2gamma}}
\end{figure}

For the decay via one-loop diagrams, the allowed range of the moduli mass is 
$m_\phi \lesssim 40$ MeV for $N=1$, as shown in Fig.~\ref{fig_2gamma}. COMPTEL and
EGRET data are taken from Ref.~\cite{Strong}, while INTEGRAL data is from Ref.~\cite{Strong2}. Diffuse spectrum above 1 MeV observed by INTEGRAL has not been
analyzed yet, but the fractional energy resolution is about 0.1\% for 
$\varepsilon_\gamma = 1-8$ MeV \cite{Roques}. We thus also show the line spectrum with
energy resolution of 0.1\% below 8 MeV. If the new analysis of INTEGRAL in the range of
$\varepsilon_\gamma = 1-8$ MeV would be done, the corresponding constraint on the moduli
mass will become $m_\phi \lesssim 10$ MeV if INTEGRAL detects the similar flux obtained by COMPTEL.

\begin{figure}[!h]
\includegraphics[width=80mm]{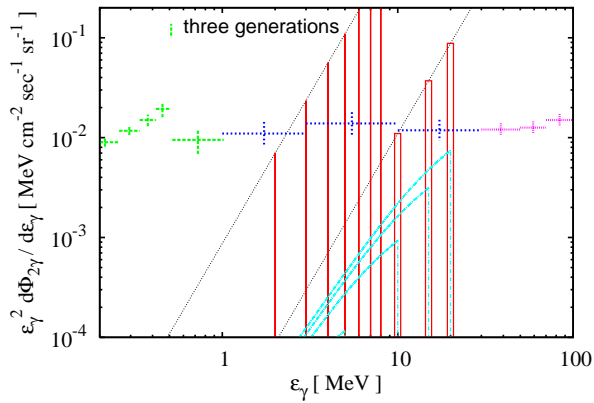}
\caption{Line spectra of two photons decayed via one-loop diagrams with $N=3$ from the 
moduli in our galaxy. Dotted lines show the fractional energy resolution of 0.08 (down) and 
0.001 (up). Data are the same as in Fig.~\ref{fig_2gamma}.
\label{fig_2gamma2}}
\end{figure}

As discussed above, the moduli is likely to couple to other leptons, if it has a coupling to
electrons (positrons) via Yukawa coupling (\ref{yukawa}). We show the emission line spectrum
for the case that the moduli couple to all three generation of leptons in Fig.~\ref{fig_2gamma2}.
Current COMPTEL data restricts the moduli mass as $m_\phi \lesssim 20$ MeV. The future 
analysis of INTEGRAL data of $\varepsilon_\gamma = 1-8$ MeV will exclude the moduli mass
above 6 MeV provided that the observed diffuse flux is as
large as that of COMPTEL. Moreover, the emission line could even be detected
for $m_\phi \lesssim 6$ MeV ($\varepsilon_\gamma \lesssim 3$ MeV). Notice that the constraint
$\varepsilon_\gamma \lesssim 3$ MeV is recently claimed in Ref.~\cite{BY}, which considered 
the inflight bremsstahlung.

\section{Conclusions}
We have considered various gamma-ray emissions from the decay of the moduli when it is 
assumed that the annihilation of the positron, into which the moduli decay, provides the
511 keV line emission observed by SPI/INTEGRAL. The flux of the photons emitted via internal 
bremsstahlung puts the limit on the moduli mass to be $m_\phi \lesssim 40$ MeV. This is model
independent constraint.

On the other hand, the moduli can decay into two photons by tree and/or one-loop diagrams, 
which gives more stringent constraint on the moduli mass. We have used extragalactic and 
galactic diffuse gamma-ray background observation to put the upper limit on the moduli mass.
As for the extragalactic background emission, the constraints are relatively loose. For instance,
the decay width of one-loop diagrams lead to $m_\phi \lesssim 60$ (30) MeV for one (three)
generation(s) of leptons running the loop. If the moduli can decay into two photons via
tree coupling, the allowed mass of the moduli is very narrow: (1.022 MeV $\leq$) 
$m_\phi \lesssim 1.5$ MeV.

Galactic diffuse gamma-ray spectrum puts the most stringent limit. If the moduli have tree level 
interactions with photons with of $O(1)$ coupling, too much photons are emitted once the flux is 
normalized to that of the 511 keV line of the positron annililations from the moduli decay. Since 
there is a coupling of the moduli with electrons (positrons), the moduli can decay into two 
photons via one-loop  diagrams. We thus obtained the mass limit of $m_\phi \lesssim 40$ MeV for the electron-positron running loop diagram. 

It is likely to have stronger coupling of the moduli with photons, since there is no symmetry to 
restrict the moduli to couple only to the first generation of lepton. We thus have the allowed mass 
range as $m_\phi \lesssim 20$ MeV currently, if all three generations of leptons run the loop.
Moreover, if the future analysis of the INTEGRAL
data above 1 MeV is done, we may even detect the line emission from the moduli due to its
high energy resolution of $\Delta\varepsilon_\gamma/\varepsilon_\gamma\simeq 0.001$, which
may become the confirmation of decaying moduli scenario.

\section*{Acknowledgments}
The authors are grateful to T.~T.~Yanagida for useful discussion.
The work of S.K. is supported by the Grant-in-Aid for Scientific Research from the
Ministry of Education, Science, Sports, and Culture of Japan, No.~17740156.

\appendix

\section{internal bremsstahlung decay rate}
Here we derive the decay rate of the internal bremsstahlung. The tree Yukawa interaction has 
the form, ${\cal L} = y\phi e \bar{e}$.  There are two diagrams contributing
to this process: the photon is emitted either electron or positron external lines. Let us consider
the process in the rest frame of the moduli, whose momentum is $q=(m_\phi, {\bf 0})$. Let the 
momenta of electron, positron, and the emitted photon be $p_1=(E_1,{\bf p_1})$, 
$p_2=(E_2,{\bf p_2})$, and $k$ with $|k|=\varepsilon_\gamma$, respectively. Then the total
invariant amplitude is calculated as
\begin{eqnarray}
\lefteqn{\overline{\left| {\cal M} \right|^2} = 
\sum_{{\rm spin},\epsilon} \left| {\cal M}_1 + {\cal M}_2 \right|^2} \nonumber \\
& & = 4y^2e^2\left[ \frac{(p_2\cdot k)}{(p_1\cdot k)} + \frac{(p_1\cdot k)}{(p_2\cdot k)} 
\right. \nonumber \\
& & \quad + \frac{2}{( p_1\cdot k)(p_2 \cdot k)}  \nonumber \\
& & \qquad\quad \times
\left\{ (p_1 \cdot p_2) + ( p_1\cdot k)   \right\} 
\left\{ (p_1 \cdot p_2) + ( p_2\cdot k)   \right\} 
\nonumber \\
& & \quad 
- \frac{m_e^2}{(p_1\cdot k)^2}\left\{ (p_1 \cdot p_2) - ( p_1\cdot k) + (p_2 \cdot k) \right\}
\nonumber \\
& & \quad 
+ \frac{3m_e^2}{(p_2\cdot k)^2}\left\{ (p_1 \cdot p_2) + ( p_1\cdot k) + (p_2 \cdot k) \right\}
\nonumber \\
& & \quad \left. + \frac{2m_e^2}{( p_1\cdot k)} +\frac{m_e^4}{( p_1\cdot k)^2}
+\frac{3m_e^4}{( p_2\cdot k)^2} - \frac{2m_e^4}{( p_1\cdot k)(p_2 \cdot k)} \right] \nonumber \\
& & = 4y^2e^2\left[ \left( \varepsilon_\gamma - m_\phi + \frac{m_\phi^2}{2\varepsilon_\gamma}
-\frac{2m_e^2}{\varepsilon_\gamma}+\frac{2m_e^2}{m_\phi}\right) \right. \nonumber \\
& &  \qquad\qquad\qquad 
\times \left\{ \frac{1}{E_1-\left(\frac{m_\phi}{2}-\varepsilon_\gamma\right)}
+\frac{1}{\frac{m_\phi}{2}-E_1}\right\} \nonumber \\
& & \qquad\qquad
+ \frac{-\frac{m_e^2}{2}+\frac{2m_e^4}{m_\phi^2}}{\left\{ E_1-\left(\frac{m_\phi}{2}-\varepsilon_\gamma\right)\right\}^2 }
+ \frac{ \frac{4m_e^2}{m_\phi}}{E_1-\left(\frac{m_\phi}{2}-\varepsilon_\gamma\right)}
\nonumber \\
& & \qquad\qquad + \left. \frac{\frac{3m_e^2}{2}}{\left( \frac{m_\phi}{2} - E_1 \right)^2} \right],
\label{inv_amp}
\end{eqnarray}
where ${\cal M}_1$ and ${\cal M}_2$ denote the invariant amplitude for each contributed diagrams. Then, the decay rate is expressed as
\begin{eqnarray}
\frac{d\Gamma_{br}}{d\varepsilon_\gamma} & = & \frac{1}{2m_\phi} \int \frac{4\pi\varepsilon_\gamma^2}{(2\pi)^3 2\varepsilon_\gamma}
\frac{d^3 p_1}{(2\pi)^3 2E_1} \frac{d^3 p_2}{(2\pi)^3 2E_2} \nonumber \\
& & \qquad  \times (2\pi)^4 \delta^{(4)}(q-p_1-p_2-k) \overline{\left| {\cal M} \right|^2} \nonumber \\
& = & \frac{1}{8(2\pi)^3} \frac{\varepsilon_\gamma}{m_\phi} \int dE_1
\overline{\left| {\cal M} \right|^2} .
\label{decayrate}
\end{eqnarray}
The range of the integration over $E_1$ is estimated as follows. Momentum conservation leads to ${\bf p_1} + {\bf p_2} + {\bf k} = {\bf 0}$, which recasted as 
$|{\bf p_2}|^2 = |{\bf p_1} + {\bf k}|^2=|{\bf p_1}|^2 +|{\bf k}|^2
 + 2 |{\bf p_1}|\varepsilon_\gamma \cos \theta$, where $\theta$ is the angle between
${\bf p_1}$ and ${\bf k}$. Since $|\cos\theta| \leq 1$, the allowed range is 
$E_- \leq E_1 \leq E_+$, where
\begin{eqnarray}
E_\pm = \frac{1}{2}\left[ m_\phi-\varepsilon_\gamma \pm \varepsilon_\gamma \left\{
1-\frac{4m_e^2}{m_\phi(m_\phi - 2\varepsilon_\gamma)}\right\}^{1/2}\right].
\end{eqnarray}
Inserting Eq.(\ref{inv_amp}) into Eq.(\ref{decayrate}), we finally obtained the decay rate as
\begin{eqnarray}
& & \frac{d\Gamma_{br} }{d\varepsilon_\gamma} = \frac{y^2 e^2}{(2\pi)^3} \nonumber \\
& & \times \left[ \left(
\frac{\varepsilon_\gamma}{m_\phi}-1+\frac{m_\phi}{2\varepsilon_\gamma}
-\frac{2m_e^2}{m_\phi \varepsilon_\gamma}+\frac{4m_e^2}{m_\phi^2}\right)
\log \frac{1+\eta}{1-\eta} \right.  \nonumber \\
& & \qquad\qquad \left. +\left( \frac{m_e^2}{m_\phi \varepsilon_\gamma} 
+\frac{2m_e^4}{m_\phi^3\varepsilon_\gamma} \right) \frac{2\eta}{1-\eta^2} \right], 
\end{eqnarray}
where 
\begin{equation}
\eta=\left(1-\frac{4m_e^2}{m_\phi(m_\phi-2\varepsilon_\gamma)}\right)^{1/2}.
\end{equation}
In the limit $m_\phi \gg m_e$, $\varepsilon_\gamma$, we get
\begin{eqnarray}
\frac{d\Gamma_{br} }{d\varepsilon_\gamma} &\simeq & 
\frac{y^2 e^2}{(2\pi)^3} \frac{m_\phi}{\varepsilon_\gamma} \log \frac{m_\phi}{m_e} 
\nonumber \\
& = & \Gamma_e \frac{4\alpha_{em}}{\pi} \frac{1}{\varepsilon_\gamma} \log \frac{m_\phi}{m_e},
\end{eqnarray}
where $\Gamma_e$ is the tree level decay rate (\ref{tree1}) or (\ref{tree2}). This represents
the familiar scaling factors, and matches to the result for annihilating particle case 
\cite{Beacom1}.



\end{document}